\documentstyle[12pt]{article}
\title{On structure of effective action in four-dimensional quantum dilaton 
       supergravity\thanks{{\it to be published in Classical and Quantum 
       Gravity}}
\begin{flushright}
                                                  {\small TSPU-TH34/96}
\end{flushright}}
\author{I.L.Buchbinder \thanks{e-mail:josephb@tspi.tomsk.su} and A.Yu.Petrov\\
      {\small\it{Department of Theoretical Physics}}\\
      {\small\it{Tomsk State Pedagogical University}}\\
      {\small\it{Tomsk 634041, Russia}}}             
\date{}
\begin{document}
\maketitle
\thispagestyle{empty}
      Short title: Effective action for dilaton supergravity
       
\newpage
\thispagestyle{empty}
\begin{abstract}
A general structure of effective action 
in new chiral superfield model associated with $N=1$, $D=4$ supergravity
is investigated. This model corresponds to finite quantum field theory and
does not demand the regularization and renormalization at effective action 
calculation. It is shown that in local approximation the effective action is 
defined by two objects called general superfield effective lagrangian and 
chiral superfield effective lagrangian. A proper-time method is generalized
for calculation of these two effective lagrangians in superfield manner.
Power expansion of the effective action in supercovariant derivatives is
formulated and the lower terms of such an expansion are calculated in explicit
superfield form.
\end{abstract}
\newpage
\setcounter{page}{1}
\section{Introduction}
In our previous paper \cite{B1} we have presented a new model of chiral
superfield and investigated its quantum formulation. This model is
conditioned by supertrace anomaly of the matter superfields in external 
superfield of $N=1$, $D=4$ supergravity. The corresponding anomaly generating 
action has been constructed in ref.[2]. We have considered a theory, action of 
which is a sum of the anomaly generating action and the action of supergravity.
Being transformed to the conformally flat superspace this action takes a form 
of action of some chiral-antichiral model in flat superspace.
 
The model under consideration possesses the remarkable properties in infrared
domain. This model is renormalizable, it is infrared free in some of couplings
and it is finite in infrared limit.
It means that the model can be considered as a natural infrared limit of quantum
supergravity with matter. 

Our goal in the present paper is to investigate a structure of the effective 
action of the above model in infrared domain. Aspects of effective action
in supersymmetric models have been studied in a number of papers [3-12].
\footnote{We would like to notice that some classical aspects of the \cite{Se3}
have also been considered in earlier paper \cite{Gates}}
The feature of our model is that the model is finite. Therefore we face
a problem of effective 
action in a finite quantum field theory and should expect 
an appearance of the new aspects.

In the recent papers [3,4,8--10] (see also the book [5]) the method of 
calculation of effective action in superfield theories has been developed and 
its applications to one- and two-loop superfield effective potentials in Wess-
Zumino model have been considered. This method is based on superfield form of
proper-time technique and provides a simple enough and manifestly supersymmetric
procedure of finding the effective action. In this paper we generalize the 
method introduced and used in refs. [3,4,8--10] in order to take into account
the features of the dilaton quantum supergravity model.

The essential feature of the model under consideration is a presence of
higher derivatives  
\footnote{We do not discuss here the question of unitarity of $S$-matrix, caused
by higher derivatives. The model under consideration should be understood as
an effective theory and
aimed for investigation of low-energy behaviour.}.
Therefore we should solve a problem of development of superfield proper-time
technique for higher derivative theories.

In any supersymmetric model containing the chiral and antichiral superfields a
corresponding effective lagrangian is represented as a sum of two contributions.
One of them depends only on chiral superfields and their space-time derivatives 
and can be called a chiral effective lagrangian. Another term depends both on
chiral and antichiral superfields and their supercovariant derivatives and can
be called a general effective lagrangian. Of course, an exact calculation
of chiral and general effective lagrangians is impossible and we have to use
the approximate schemes. 
It is natural to search the effective lagrangians in form
of power expansion in supercovariant derivatives. As we will see the remarkable 
properties of the model under consideration stipulated by its finiteness allow 
to write down each term in these expansions in explicit form up to some constants.
In particular, here we can not expect in the effective lagrangians any logarithmic
contributions which are most typical in the standard quantum field theories
with divergences. As a result we have the only problem how to evaluate above
constants.

Use of the superfield proper-time technique allows to calculate any term in 
above expansions for one-loop general and chiral effective lagrangians. In this
paper we find the lower terms in these expansions and show they have no any
divergences and 
their structure corresponds to result of general analysis of the 
effective lagrangian on the base of the dimension and finiteness of the theory.

The paper is organized as follows. The section 2 is devoted to description of 
the model, definition of superfield effective actions and its symmetries and
investigating the properties of the general and chiral lagrangians.
In section 3 the method of calculating the one-loop effective action is 
developed. In particular we show that the one-loop effective action for chiral-
antichiral superfield model can be expressed in terms of some vector multiplet
model. The sections 4 and 5 are devoted to calculating the lower terms in
expansions of general and chiral one-loop lagrangians in derivatives.
In section 6 we show the correspondense between the model under consideration 
and massless Wess-Zumino model. The summary is devoted to general discussion of
the results.

\section{Structure of effective action}
The previous paper \cite {B1} has been devoted to investigation of structure of
renormalization and infrared asymptotical behaviour of four-dimensional
dilaton supergravity model. The action of this model is given as a sum of the 
action generating the superconformal anomaly and the classical action of $N=1$
supergravity. The complete action in the conformally flat superspace has the 
form
\begin {eqnarray}
\label{action}
S&=&\int d^8z
    (-\frac{Q^2}{2{(4\pi)}^2}\bar{\sigma}\Box\sigma+
\bar{D}^{\dot\alpha}\bar{\sigma}D^{\alpha}\sigma\times
({\xi_1\partial_{\alpha}}_{\dot{\alpha}}(\sigma+\bar{\sigma})+\nonumber\\
&+&\xi_2\bar{D}_{\dot{\alpha}}\bar{\sigma} D_{\alpha}\sigma)-
\frac{m^2}{2} e^{\sigma+\bar{\sigma}})+(\Lambda\int d^6z e^{3\sigma}+h.c.)
\end {eqnarray}
Here $\sigma=\ln \Phi$ where $\Phi$ is a chiral supergravity compensator
(the only dynamical field in this theory),  $Q^2$, $\xi_1$, $\xi_2$, $m^2$,
$\Lambda$ are couplings. $\sigma$ is dimensionless.

We have investigated renormalization group equations for the effective
couplings of this theory and proved that in infrared limit (i.e at
$t\rightarrow -\infty$) they satisfy the conditions:
$\xi_1(t)=\xi_2(t)=m^2(t)=0$, $Q^2(t)=Q^2\neq 0$.
So, action of the theory in infrared domain can be written in the form
\begin{equation}
\label{iract}
S=\int d^8 z \big( -\frac{1}{2}\frac{Q^2}{{(4\pi)}^2}\bar{\sigma}\Box\sigma\big)
+\big(\Lambda\int d^6 z e^{3\sigma}+ h.c.\big)
\end{equation}

It is easy to show that classical action (\ref{iract}) is invariant under 
transformations
\begin{eqnarray}
\label{inv}
\sigma\to \sigma+\alpha,\ \bar{\sigma}\to \bar{\sigma}+\beta\\
\Lambda\to e^{-3\alpha}\Lambda,\ \bar{\Lambda}\to e^{-3\beta}\Lambda
\nonumber
\end{eqnarray}
$\alpha$ and $\beta$ are the independent parameters,
for example, for chiral effective action independent on $\bar{\sigma}$ we put
$\beta=0$ and $\Lambda$ is not transformed. 
After introduction of classical background field $\sigma$ and the quantum one 
$\chi$ \cite{BO} we arrive at the following quantum action for $\chi$ which will
be also invariant under transformations (\ref{inv}) of background 
superfields $\sigma$ and $\bar{\sigma}$.
\begin{equation}
\label{qs}
S_q=\int d^8 z \big( -\frac{Q^2}{2{(4\pi)}^2}\bar{\chi}\Box\chi\big)+
\big(\Lambda\int d^6 z e^{3\sigma}(e^{3\chi}-1-3\chi)+ h.c.\big)
\end{equation}
The quantum field $\chi$ in (4) is not transformed under these transformations
(\ref{inv}).

The effective action of the theory has the form
\begin{equation}
\Gamma[\sigma,\bar{\sigma}]=S[\sigma,\bar{\sigma}]+
\bar{\Gamma}[\sigma,\bar{\sigma}]
\end{equation}
where
$S[\sigma,\bar{\sigma}]$ is the classical action (\ref{iract}) and
$\bar{\Gamma}[\sigma,\bar{\sigma}]$ is the sum of all quantum corrections.
Here the functional
$\bar{\Gamma}[\sigma,\bar{\sigma}]$ is defined as follows (see f.e. \cite{BO})
\begin{equation}
\exp(i\bar{\Gamma}[\sigma,\bar{\sigma}])=\int D\chi D\bar{\chi} exp(iS_q
[\sigma, \bar{\sigma};\chi, \bar{\chi}])
\end{equation}
It was proved in 
the previous paper \cite{B1} that the theory under consideration
is finite so any anomalies in this theory cannot arise. Therefore we should 
expect
that the effective action is also invariant under the transformations
(\ref{inv}).

We suppose that the quantum correction to the effective action 
can be rewritten as
\begin{equation}
\bar{\Gamma}=\int d^8 z L+(\int d^6 z {\cal L}_c+h.c.)
\end{equation}
Here $L$ is a general effective lagrangian,
${\cal L}_c$ is a chiral effective lagrangian satisfying the condition
$\bar{D}_{\dot{\alpha}}{\cal L}_c=0$ i.e. ${\cal L}_c$
is a chiral superfield itself. 
${\cal L}_c (\sigma)$ 
depends on $\sigma$ and its space-time derivatives but not 
on $D_{\alpha}\sigma$, $D^2\sigma$.
One can suppose that ${\cal L}_c (\sigma)$ can depend on 
$\bar{D}^2\bar{\sigma}$ since $\bar{D}^2\bar{\sigma}$ is a chiral superfield, 
i.e. in the expansion of ${\cal L}_c (\sigma)$ the terms of the form 
$f(\sigma, \partial\sigma, \partial^2\sigma\ldots){(\bar{D}^2\bar{\sigma})}^n$ 
formally can arise but their contribution to the effective action of the form
$\int d^6 z 
f(\sigma, \partial\sigma, \partial^2\sigma\ldots){(\bar{D}^2\bar{\sigma})}^n$
can be rewritten as an integral over whole superspace as
$\int d^8 z 
f(\sigma, \partial\sigma, \partial^2\sigma\ldots){(\bar{D}^2
\bar{\sigma})}^{n-1}$
It means we can consider such terms as the corrections to the general effective 
lagrangian and put ${\cal L}_c (\sigma)$ to be independent on $\bar{\sigma}$.

The effective lagrangians $L(\sigma,\bar{\sigma})$ and ${\cal L}_c(\sigma)$ can 
be represented in the form of power expansion in derivatives of $\sigma$, 
$\bar{\sigma}$. 
The lower correction to $L$ should has the form 
$L \sim \Lambda^{k_1} \bar{\Lambda}^{k_2} e^{n(\sigma+\bar{\sigma})}$ where
$k_1, k_2$ and $n$ are some numbers. 
It is turned out that all these numbers, in principle,
can be found exactly without any calculations. Namely, requirements that
the dimension of $L$ is 2 and that the action is invariant under the 
transformations (\ref{inv}) allow to obtain the lower contribution to 
general effective lagrangian in the form
\begin{equation}
K= \rho {(\Lambda\bar{\Lambda})}^{1/3} e^{\sigma+\bar{\sigma}}
\end{equation}
where $\rho$ is some dimensionless constant which 
can be found only at straightforward calculations.

The lower contribution to ${\cal L}_c$ should have the form of a linear 
combination of the terms\\
$\Lambda^{q_1}\bar{\Lambda}^{q_2}e^{p\sigma}
\partial_m\sigma\partial^m\sigma$ and 
$\Lambda^{q_1}\bar{\Lambda}^{q_2}e^{p\sigma}\Box\sigma$ 
where $q_1$, $q_2$, $p$ are some numbers. The dimension of ${\cal L}_c$
is 3, hence $3(q_1+q_2)+2=3$. The invariance of ${\cal L}_c$ under above 
transformations leads to $-3q_1+p=0$ (we use transformations at $\beta=0$ 
there). As a result $q_1=p/3$, $q_2=1/3-p/3$. Hence the lower contribution
to ${\cal L}_c$ looks like this
\begin{equation}
{\cal L}_c=\bar{\Lambda}^{1/3}\big[ 
\sum_{p_1} {(\frac{\Lambda}{\bar{\Lambda}})}^{p_1} \rho_{p_1} e^{p_1\sigma}
\partial_m\sigma\partial^m\sigma+ \sum_{p_2} {(\frac{\Lambda}
{\bar{\Lambda}})}^{p_2} \rho_{p_2} e^{p_2\sigma}\Box\sigma\big]
\end{equation}
where $\rho_{p_1}$ and $\rho_{p_2}$ are some dimensionless constants. These
constants and the values of $p_1$ and $p_2$ can be found at straightforward 
calculations.

In this paper we will calculate the lower corrections to the general effective 
lagrangian $L$ and to chiral effective lagrangian ${\cal L}_c(\sigma)$ at the 
one-loop level exactly.

\section{Problems of calculating the one-loop effective action}

It is well known that the one-loop correction to the effective action 
$\bar{\Gamma}^{(1)}$ has the form
\begin{equation}
\exp(i\bar{\Gamma}^{(1)})=\int D\chi D\bar{\chi}
\exp(iS^{(2)}_q[\sigma, \bar{\sigma};\chi, \bar{\chi}])
\end{equation}
where $S^{(2)}_q$ is a quadratic part of $S_q$ (\ref{qs}).
In the case under consideration one can obtain
\begin{eqnarray}
\label{ef}
\exp(i\bar{\Gamma}^{(1)})&=&\int D\chi D\bar{\chi} \exp\big\{\big[
\int d^8 z \big( -\frac{1}{2}\frac{Q^2}{{(4\pi)}^2}\bar{\chi}\Box\chi\big)+\\
&+&\big(\Lambda\int d^6 z e^{3\sigma}\frac{{(3\chi)}^2}{2}+
 h.c.\big)\big]\big\}=Det^{-1/2} {\cal H}\nonumber
\end{eqnarray}
where
$${\cal H}=\left(
\begin{array}{cc}
 9 \Lambda e^{3\sigma} & \Box \frac{\bar{D}^2}{4}\\
\Box \frac{D^2}{4} & 9 \bar{\Lambda} e^{3\bar{\sigma}}
\end{array}
\right)$$
Straightforward calculating Det ${\cal H}$ is a very complicated problem since 
elements of this matrix are defined in different subspaces of the superspace and
mix the chiralities. In the 
papers [3,4] where the Wess-Zumino model was investigated a 
trick allowing to avoid these difficulties has been suggested and
the effective action has been expressed in terms of Green function of 
the real scalar 
superfield. We will show that the trick carried out in these papers can be 
applied to the theory under consideration too.

We consider a theory of a real scalar superfield with the action
\begin{equation}
\label{act1}
S=-\frac{Q^2}{{(4\pi)}^2}\frac{1}{16}\int d^8 z V D^{\alpha}\bar{D}^2
D_{\alpha} \Box V
\end{equation}
and calculate a corresponding one-loop effective action.

This theory is invariant under gauge transformations
\begin{equation}
\label{trans}
V\rightarrow V^{\Psi}=V+i(\bar{\Psi}-\Psi),
\bar{D}_{\dot{\alpha}}\Psi=0, D_{\alpha}\bar{\Psi}=0
\end{equation}
We choose the gauge-fixing functions as follows
\begin{eqnarray}
F(V)=\frac{1}{4}\bar{D}^2 V-\chi\\
\bar{F}(V)=\frac{1}{4}D^2 V-\bar{\chi}
\end{eqnarray}
Their variation under transformations (\ref{trans}) has the form
\begin{equation}
\delta\left(\begin{array}{c}
F(V)\\
\bar{F}(V)
\end{array}\right)=
-M^{(0)}\left(\begin{array}{c}
i\Psi\\
i\bar{\Psi}
\end{array}\right)
\end{equation}
where
\begin{equation}
M^{(0)}=
\left(\begin{array}{cc}
0 & -\frac{1}{4}\bar{D}^2\\
-\frac{1}{4} D^2 & 0
\end{array}\right)
\nonumber
\end{equation}
As a result we can write

\begin{equation}
det M^{(0)}\int D\Psi D\bar{\Psi} \delta_+ (F(V^{\Psi}))
\delta_- (\bar{F}(V^{\Psi}))=1
\end{equation}

The one-loop effective action $W_v$ for the theory (\ref{act1}) has the form
\begin{eqnarray}
\label{Wv}
\exp(iW_v)&=&\int DV det M^{(0)} \delta_+ (F(V))\delta_- (\bar{F}(V))
\exp \{-\frac{1}{16}\frac{Q^2}{{(4\pi)}^2}\times\\
&\times&\int d^8 z V D^{\alpha}\bar{D}^2
D_{\alpha} \Box V\}\nonumber
\end{eqnarray}
By definition $\exp(iW_v)$ is a constant. So, after multiplying (\ref{ef}) and
(\ref{Wv}) we obtain
\begin{eqnarray}
\exp(i\bar{\Gamma}^{(1)}+iW_v)&=&\int D\chi D\bar{\chi} DV
det M^{(0)} \delta_+ (\frac{1}{4}\bar{D}^2 V-\chi)
\delta_- (\frac{1}{4}\bar{D}^2 V-\chi)\times\nonumber\\
&\times&\exp \{-\frac{1}{16}\frac{Q^2}{{(4\pi)}^2}\int d^8 z
V D^{\alpha}\bar{D}^2 D_{\alpha} \Box V\}-\\
&-&\big\{\big[\int d^8 z \big( -\frac{1}{2}\frac{Q^2}{{(4\pi)}^2}
\bar{\chi}\Box\chi\big)+
\big(\Lambda\int d^6 z e^{3\sigma}\frac{{(3\chi)}^2}{2}+
 h.c.\big)\big]\big\}\nonumber
\end{eqnarray}
Let us integrate over $\chi$ using delta-functions, then one obtains
\begin{eqnarray}
& &\exp(i\bar{\Gamma}^{(1)})\exp(iW_v) det^{-1} M^{(0)}=\nonumber\\
&=&DV\exp \{\frac{1}{2}\int d^8 z\frac{Q^2}{{(4\pi)}^2}
V \Box^2 V+\big(\frac{9\Lambda}{2}
\int d^6 z e^{3\sigma}{(\frac{\bar{D}^2 V}{4})}^2+\\
&+&\frac{9\bar{\Lambda}}{2}\int d^6 z e^{3\bar{\sigma}}{(\frac{D^2 V}{4})}^2
\big)\}=\int DV \exp(\frac{i}{2} V\Delta V)\nonumber
\end{eqnarray}
Since $W_v$ and $\det M^{(0)}$ are constants by the definitions we can
obtain one-loop effective action for the theory under consideration in the form
\begin{equation}
\label{expr}
\bar{\Gamma}^{(1)}=\frac{i}{2} Tr ln \Delta
\end{equation}
where
\begin{equation}
\Delta=\frac{Q^2}{{(4\pi)}^2}\Box^2-9\Lambda e^{3\sigma}\frac{\bar{D}^2}{4}-
9\bar{\Lambda} e^{3\bar{\sigma}}\frac{D^2}{4}
\end{equation}
We would like to notice that the operator $\Delta$ contains the higher 
derivatives (it contains the four derivatives being written in component form).

From (\ref{expr}) we obtain
\begin{equation}
\bar{\Gamma}^{(1)}=-\frac{i}{2}\int_0^{\infty} \frac{ds}{s} Tr e^{is\Delta}
\end{equation}
Our aim is to calculate $\Omega(\sigma,\bar{\sigma}|s)=e^{is\Delta}$.
Now let us consider general effective lagrangian and chiral effective lagrangian
separately.

\section{General effective lagrangian}
We start with calculation of the lower correction to general effective 
lagrangian, i.e. a case 
when the fields $\sigma$, $\bar{\sigma}$ and hence $\Omega(\sigma,\bar{\sigma}|s)$ do 
not depend on derivatives of the fields. 
It is easy to see that the $\Omega$ has the form
\begin{eqnarray}
\Omega(\sigma,\bar{\sigma}|s)&=&\exp (is(\frac{Q^2}{{(4\pi)}^2}\Box^2
-\frac{9}{4}\Lambda e^{3\sigma}\bar{D}^2-\frac{9}{4}\bar{\Lambda}
e^{3\bar{\sigma}}D^2))=\\
&=&\exp(is(-\frac{9}{4}\Lambda e^{3\sigma}\bar{D}^2-\frac{9}{4}\bar{\Lambda}
e^{3\bar{\sigma}}D^2))
\exp(is\frac{Q^2}{{(4\pi)}^2}\Box^2)\nonumber
\end{eqnarray}

Hence, to calculate a trace of $\Omega$ we must compute:
$$1.\ \tilde{\Omega}=\exp(is(-\frac{9}{4}\Lambda e^{3\sigma}\bar{D}^2-
\frac{9}{4}\bar{\Lambda}e^{3\bar{\sigma}}D^2))$$
$$2.\ U(x,x'|s)=\exp(is\frac{Q^2}{{(4\pi)}^2}\Box^2)\delta^4(x-x')$$

For simplifying calculations we will introduce a new parameter $s$ using a rule
$s\frac{Q^2}{{(4\pi)}^2}\to s$
and denote 
$$h=-\frac{9}{4}\Lambda\frac{{(4\pi)}^2}{Q^2},\
\bar{h}=-\frac{9}{4}\bar{\Lambda}\frac{{(4\pi)}^2}{Q^2},\
\varphi=e^{3\sigma},\ \bar{\varphi}=e^{3\bar{\sigma}}$$
Then $\Omega$ has the form

\begin{equation}
\Omega(\sigma,\bar{\sigma}|s)=\exp (is(\Box^2+h
\varphi\bar{D}^2+\bar{h}\bar{\varphi}D^2))=
\exp(is(h\varphi\bar{D}^2+\bar{h}\bar{\varphi}D^2))\exp(is\Box^2)
\end{equation}

Let us consider at first the function $\tilde{\Omega}=
\exp(is(h\varphi\bar{D}^2+\bar{h}\bar{\varphi}D^2))$
satisfying the equation
\begin{equation}
i\frac{\partial\tilde{\Omega}}{\partial s}=-\tilde{\Omega}\tilde{\Delta}
\end{equation}
Here 
$$ \tilde{\Delta}=h\varphi\bar{D}^2+
\bar{h}\bar{\varphi}D^2,\ \tilde{\Omega}|_{s=0}=1$$
Let us represent $\tilde{\Omega}$ in the form of expansion in spinor 
supercovariant derivatives \cite{BK2}, \cite{BK3}:
\begin{eqnarray}
\label{omega}
\tilde{\Omega}= 1+\frac{1}{16}A(\tilde{s})\bar{D}^2 D^2+\frac{1}{16}\tilde{A}
(\tilde{s}) D^2 \bar{D}^2+\frac{1}{8} B^{\alpha} (\tilde{s}) D_{\alpha}
\bar{D}^2+\tilde{B}_{\dot{\alpha}}(\tilde{s})\bar{D}^{\dot{\alpha}} D^2+\\
\frac{1}{4}C(\tilde{s})D^2+\frac{1}{4}\tilde{C}(\tilde{s})\bar{D}^2\nonumber
\end{eqnarray}
The solution of the eq. (27) for $\tilde{\Omega}$ (28) with the
initial conditions: $A,\tilde{A},B,\tilde{B},C,\tilde{C}$ at $s=0$
are equal to zero looks like this
\begin{equation}
\label{coef}
A=\tilde{A}=\frac{1}{\Box}[\cosh(4\tilde{s}
\sqrt{h\bar{h}\varphi\bar{\varphi}\Box})-1]
\end{equation}
where $\tilde{s}=-is$
(see some details in refs. [3,4]).
It follows from (24, \ref{omega}) that only terms containing $A$ and $\tilde{A}$
give non-zero contribution when calculating trace of $\Omega$.
The eq. (\ref{coef}) shows that lower contribution to one-loop general effective
lagrangian
depends only on $\varphi\bar{\varphi}=e^{3(\sigma+\bar{\sigma})}$.

After calculations described in Appendix A we obtain the lower contribution to 
one-loop general effective lagrangian in the form
\begin{equation}
K^{(1)}= \Big(\frac{c}{Q^{4/3}}\Big)
{(\Lambda\bar{\Lambda})}^{1/3} e^{\sigma+\bar{\sigma}}
\end{equation}
where $c$ is a finite constant. The explicit form for $c$ is given in Appendix 
A.

This result coincides with the general structure of $K$ obtained in the section
2. We also see that although there is no the 'massive' term like $e^{\sigma+
\bar{\sigma}}$ in this theory at a classical level such a term arises as a 
quantum correction.
We can conclude that the lower contribution to one-loop general effective 
lagrangian is finite
in correspondence with the results of the previous paper \cite{B1}.

\section{Chiral effective lagrangian}
In this section we consider calculating the lower contribution to one-loop 
chiral effective
lagrangian ${\cal L}^{(1)}_c$ i.e the part of one-loop effective action which
is a chiral superfield itself and does not depend on $\bar{\sigma}$.
To calculate the chiral effective lagrangian in the model under consideration 
one should put $\bar{\sigma}=0$ in the operator $\Omega(\sigma,\bar{\sigma})$ 
introduced in the section 3. Here and further we will denote
$\Omega(\sigma)=\Omega(\sigma,\bar{\sigma})|_{\bar{\sigma}=0}$.
The operator $\Omega(\sigma)$ has the form
\begin{equation}
\Omega(\sigma)=e^{is\Delta}=
\exp (is(\Box^2+h\varphi\bar{D}^2+\bar{h} D^2))
\end{equation}
The part of one-loop effective action depending only on the chiral superfield
$\varphi$ can be written as
\begin{equation}
\label{gamma}
\tilde{\Gamma}^{(1)}=-\frac{i}{2}\int_0^{\infty} \frac{ds}{s}Tr e^{is\Delta}
\end{equation}
where
$\Delta=\Box^2+h\varphi\bar{D}^2+\bar{h} D^2$.
We will call the one-loop chiral effective lagrangian ${\cal L}^{(1)}_c$ that 
part of $\tilde{\Gamma}^{(1)}$ which
is a chiral superfield itself and depends only on $\varphi$ and its space-time 
derivatives.

We can represent ${\cal L}^{(1)}_c$ as a power expansion in space-time 
derivatives. Our aim here consists in calculating the lower terms in this 
expansion.

Let us consider $e^{is\Delta}$ in details. We can write it in the form
\begin{equation}
\label{haus}
e^{is\Delta}=e^{is\Box^2}e^{is(h\varphi\bar{D}^2+\bar{h} D^2)}M
\end{equation}
where $M$ depends on different commutators of the form
$$[\Box^2,[\ldots[h\varphi\bar{D}^2+\bar{h} D^2,[\ldots
[\Box^2,h\varphi\bar{D}^2+\bar{h} D^2]]]]$$
and can be represented in the form
$M=1+(commutators)$ since if all these commutators are equal to zero, $M$ 
will be evidently equal to 1.
Due to the cyclic property of trace one can write
$$Tr e^{is\Delta}=Tr e^{is(h\varphi\bar{D}^2+\bar{h} D^2)}M e^{is\Box^2}$$
Further, it should be noted that the lower commutator in $M$ of the form
$[\Box^2,h\varphi\bar{D}^2+\bar{h} D^2]$, and hence higher commutators contained
in $M$ depend at 
least on first space-time derivative of $\varphi$ so $M$ can be 
written in the form:
$$M=1+\partial_m\varphi a^m+\ldots$$
where $a_m$ depends on 
$\varphi$, $D_{\alpha}\varphi$, $D^2\varphi$ and contains 
$\partial$ in 
different powers, dots denote all terms proportional to second and higher space-
time derivatives of $\varphi$.

Then, after calculation $\tilde{\Omega}=
e^{is(h\varphi\bar{D}^2+\bar{h} D^2)}$ we will obtain the coefficient
$A$ in the expansion (28) in the form
$$A(s)=A_1(s|\varphi, \partial\varphi..)+D^{\alpha}\varphi D_{\alpha}\varphi
A_2(s|\varphi, \partial\varphi..)
+D^2\varphi A_3(s|\varphi, \partial\varphi..)$$
and $\tilde{A}$ in the analogous form. Here dots denote higher space-time
derivatives of $\varphi$ and powers of $\Box$.
Because of (\ref{kahl}) the one-loop chiral effective lagrangian has the form
$${\cal L}^{(1)}_c=-\frac{i}{2}(-\frac{1}{4})\int_0^{\infty}\frac{ds}{s}
\bar{D}^2<A(\tilde{s})+\tilde{A}(\tilde{s})>U(x,x';s)|_{x=x'}$$
which can be rewritten as follows
\begin{eqnarray}
{\cal L}^{(1)}_c&=&2i\int_0^{\infty}\frac{ds}{s}
\Big(\partial_m\varphi\partial^m\varphi
\big(A_2(s|\varphi, \partial\varphi..)+
\tilde{A}_2(s|\varphi, \partial\varphi..)\big)
+\nonumber\\
&+&\Box\varphi\big(A_3(s|\varphi, \partial\varphi..)+
\tilde{A}_3(s|\varphi, \partial\varphi..)\big)
\Big)U(x,x';s)|_{x=x'}\nonumber
\end{eqnarray}
So dependence of $A$ and $\tilde{A}$ on $\partial\varphi$ and higher
space-time derivatives of $\varphi$ gives contribution only to 
third and higher orders in expansion of ${\cal L}^{(1)}_c$ in derivatives.
It is evidently that taking into account the commutators containing in $M$ 
we will get the higher orders in derivatives by analogous reasons. 
Hence to calculate chiral effective action up to the second order in derivatives
one should:\\
1. put $M$ in (\ref{haus}) to be equal to 1, i.e. put
$$e^{is\Delta}=e^{is\Box^2}e^{is(h\varphi\bar{D}^2+\bar{h} D^2)}$$\\
2.When calculating $e^{is(h\varphi\bar{D}^2+\bar{h} D^2)}$ put
$\partial_m\varphi$ and higher space-time derivatives of $\varphi$ to be equal 
to zero.

So we can represent $\Omega(\sigma)$ for calculating the chiral effective
action in the form 
$$\Omega(\sigma)=\tilde{\Omega}(\sigma)e^{is\Box^2}$$
where $\tilde{\Omega}(\sigma)=\exp (is(h\varphi\bar{D}^2+\bar{h} D^2))=
\exp(is\tilde{\Delta})$. Here $\tilde{\Omega}(\sigma)$ satisfies the equation:
$$i\frac{\partial \tilde{\Omega}(\sigma)}{\partial s}=-\tilde{\Omega}(\sigma)
(h\varphi\bar{D}^2+h D^2)$$
$\tilde{\Omega}$ can be represented in the form (\ref{omega}).
The initial conditions coincide with those one for the general effective
potential, namely, $\tilde{\Omega}$=1 and $A,\tilde{A}$, $B,\tilde{B}$, 
$C,\tilde{C}$ all are equal to zero at $s=0$.

We can solve the system for $\tilde{\Omega}$ by the method analogous to one
of refs. [3,4].
The operatoral coefficients $A,\tilde{A}$ necessary for our calculations have
the form
\begin{eqnarray}
\label{cf}
\tilde{A}&=&\frac{1}{\Box}(\cosh(W\tilde{s})-1)\\
A&=&-16\bar{h}\big\{-8h\varphi\tilde{s}^2+\frac{64h\bar{h}}{W^2}\big(
\frac{\sinh(W\tilde{s})}{W}-\tilde{s}\big)D^2\varphi+\nonumber\\
&+&2^{13}\frac{h^2\bar{h}^2}{W^5}D^{\alpha}\varphi D_{\alpha}\varphi[W\tilde{s}
\cosh(W\tilde{s})-3\sinh(W\tilde{s})+2W\tilde{s}]\big\}\nonumber
\end{eqnarray}
where $W=16\sqrt{h\bar{h}\varphi\Box}$ and $\varphi=e^{3\sigma}$.

After the calculations described in Appendix B one obtains the one-loop chiral
effective action up to the second order in derivatives in the form (\ref{res1}).
Restoring $\sigma$ in this expression using the definition $\varphi=e^{3\sigma}$
we arrive at ${\cal L}^{(1)}_c$ in the form
\begin{eqnarray}
\label{chir}
{\cal L}^{(1)}_c&=&\bar{\Lambda}^{1/3}
\{((c_1+3c_3) {\big(\frac{\bar{\Lambda}}{\Lambda}\big)}^{1/3} e^{-\sigma} +
c_2 {\big(\frac{\bar{\Lambda}}{\Lambda}\big)}^{-2/3}e^{2\sigma} +3c_4 
{\big(\frac{\bar{\Lambda}}{\Lambda}\big)}^{4/3}e^{-4\sigma})\times\\
&\times&\partial^m\sigma \partial_m\sigma+
\Box\sigma (c_3 {\big(\frac{\bar{\Lambda}}{\Lambda}\big)}^{1/3}e^{-\sigma} +
c_4 {\big(\frac{\bar{\Lambda}}{\Lambda}\big)}^{4/3}e^{-4\sigma})\}\nonumber
\end{eqnarray}
where $c_1$, $c_2$, $c_3$, $c_4$ are finite constants given in Appendix B.

The eq. (35) is the final result for the lower correction to the one-loop chiral
effective action. We see that it is not equal to zero but essentially
dependent on space-time derivatives. This result corresponds to the general
structure of ${\cal L}^{(1)}_c$ obtained in the section 2.
We also can conclude that it is finite in correspondence with the results
of the previous paper \cite{B1}.

\section{Generating of Wess-Zumino model}
We have calculated the lower correction to the one-loop effective action for
the theory under consideration and proved that the effective action in this 
approximation has the form:
\begin{eqnarray}
\Gamma&=&\int d^8 z \big( -\frac{Q^2}{{(4\pi)}^2}\frac{1}{2}\bar{\sigma}\Box
\sigma\big)+
\big(\Lambda\int d^6 z e^{3\sigma}+ h.c.\big)+\\
&+&c {\Big(\frac{\Lambda}{Q^2}\Big)}^{2/3}
\int d^8 z e^{\sigma+\bar{\sigma}}+\nonumber\\
&+&{\Big(\frac{\Lambda}{Q^2}\Big)}^{1/3}[\int d^6 z
\{( (c_1+3c_3) e^{-\sigma} +c_2 e^{2\sigma} +3c_4 e^{-4\sigma})
\partial^m\sigma \partial_m\sigma+\nonumber\\
&+&\Box\sigma (c_3 e^{-\sigma} +c_4 e^{-4\sigma})\}+h.c.]\nonumber
\end{eqnarray}
Here ones put $\bar{\Lambda}=\Lambda$.
If we also put $\partial_m\sigma=0$, $\partial_m\bar{\sigma}=0$ (i.e. consider
the superfields slowly varying in space-time) we will obtain the effective 
action in the form
\begin{equation}
\Gamma=c {\Big(\frac{\Lambda}{Q^2}\Big)}^{2/3}
\int d^8 z e^{\sigma+\bar{\sigma}}+
\big(\Lambda\int d^6 z e^{3\sigma}+ h.c.\big)
\end{equation}

Let us denote 
$\phi=\sqrt{c}{\Big(\frac{\Lambda}{Q^2}\Big)}^{1/3}e^{\sigma}$,  
$\bar{\phi}=\sqrt{c}{\Big(\frac{\Lambda}{Q^2}\Big)}^{1/3}e^{\bar{\sigma}}$.
Then the eq. (37) can be rewritten as follows
\begin{equation}
\label{WZ}
\Gamma=\int d^8 z \phi\bar{\phi}+
\big(\lambda \int d^6 z {\phi}^3+ h.c.)
\end{equation}
where $\lambda=Q^2 c^{-3/2}$.
This action describes dynamics of a massless chiral superfield $\phi$ of 
dimension equal to 1 and completely coincides with the action of massless Wess-
Zumino model. So one can conclude that
the four-dimensional model of quantum supergravity chiral compensator in 
infrared limit leads to the Wess-Zumino model.

We see also that $\Lambda$ is absent in the action (\ref{WZ}), its role
consists in transformation of dimensionless field $\Phi$ ($\Phi=e^{\sigma}$ is 
a chiral compensator) to the field $\phi$ with dimension 1.

{\section{Summary}}
We have investigated a structure of effective action for the dilaton 
supergravity in the infrared domain where the model is finite. The effective 
action is defined by 
general effective 
lagrangian and chiral 
effective lagrangian. We have considered a generic form of lower 
contributions to these
effective lagrangians within expansions in supercovariant derivatives and found 
that the finiteness of the model allows to write down the terms in above 
expansions up to some constants.

Generalized superfield proper-time technique has been developed and its 
application to calculating the one-loop effective lagrangians has been 
investigated. In particular, the lower contributions to the 
general effective lagrangian and chiral effective lagrangian have been 
calculated in explicit superfield form.

We have studied a structure of one-loop effective action constructed on the 
base of the lower contributions to the effective lagrangians for the
superfields slowly varying in space-time. It is shown that in this limit the 
effective action is reduced to the action of standard massless Wess-Zumino 
model.

We would like to notice in conclusion that the non-supersymmetric model of 
dilaton gravity has been developed in ref. [14]. The structure of effective 
action in this model was considered in refs. [15-22]. The supersymmetric model 
studied here is finite unlike non-supersymmetric version. It leads to simple
enough structure of the effective lagrangians in compare with one in non-
supersymmetric model [15-22].

\vspace{5mm}
{\Large\bf{Acknowledgements}}

The authors are grateful to S.D.Odintsov and S.M.Kuzenko for interesting 
discussions on some aspects of the work. We thankful to S.J.Gates paying our
attention to paper [23]. The work was supported in part by Russian Foundation 
for Basic Research under the project No. 94-02-03234.
I.L.B. thanks D.Luest, D.Ebert and
H.Dorn for their hospitality during his visit to Institute of Physics, Humboldt
Berlin University where most part of the work has been fulfilled.
This visit was supported by Deutsche Forschungsgemeinschaft under
contract DFG$-$436 RUS 113.

\newpage
{\large\bf{Appendix A. Calculation of one-loop general effective potential}}

\vspace{2mm}

\setcounter{equation}{0}
\renewcommand{\theequation}{A.\Roman{equation}}
It is well known from refs. [3,4] that the one-loop effective action
has the form
\begin{equation}
\label{kahl}
\bar{\Gamma}^{(1)}=-\frac{i}{2}\int d^8 z \int_0^{\infty}\frac{ds}{s}
<A(\tilde{s})+\tilde{A}(\tilde{s})>U(x,x';s)|_{x=x'}
\end{equation}
where $A(\tilde{s})$ and $\tilde{A}(\tilde{s})$ are given by (28).
We will calculate 
$$<A(\tilde{s})+\tilde{A}(\tilde{s})>U(x,x';s)|_{x=x'}$$
One can represent this expression in the form
$$<A(\tilde{s})+\tilde{A}(\tilde{s})>U(x,x';s)=\sum_{n=0}^{\infty}
f_n(\tilde{s})\Box^n U(x,x';s)$$
Here $U(x,x';s)=\exp(i(s+i\epsilon)\Box^2)\delta^4(x-x')$.
Our aim is 
to compute $\Box^n U(x,x;s)|_{x=x'}$. Using the Fourier representation
$$U(x,x';s)=\int \frac{d^4 k}{{(2\pi)}^4}e^{i(s+i\epsilon) k^4-ik(x-x')},$$
ones obtain
$$\Box^n U(x,x';s)=\int \frac{d^4 k}{{(2\pi)}^4}{(-k^2)}^n
e^{i(s+i\epsilon)k^4-ik(x-x')}$$
When $x=x'$ we find
$$\Box^n U(x,x';s)|_{x=x'}=\int \frac{d^4 k}{{(2\pi)}^4}{(-k^2)}^n
e^{i(s+i\epsilon) k^4}$$

If $n$ is even, $n=2l$, then
$$\Box^n U(x,x;s)={(-i\frac{\partial}{\partial s})}^l
\int \frac{d^4 k}{{(2\pi)}^4} e^{i(s+i\epsilon)k^4}$$
if $n=2l+1$ then
$$\Box^n U(x,x;s)={(-i\frac{\partial}{\partial s})}^l
\int \frac{d^4 k}{{(2\pi)}^4}(-k^2) e^{i(s+i\epsilon)k^4}$$
where we take into account $$\Box^2 U=-i\frac{\partial}{\partial s}U$$
For the integrals arisen one obtains
\begin{eqnarray}
\label{ints}
\int \frac{d^4 k}{{(2\pi)}^4} e^{i(s+i\epsilon)k^4}=\frac{1}{32\pi^2\tilde{s}}\\
\int \frac{d^4 k}{{(2\pi)}^4}(-k^2) e^{i(s+i\epsilon)k^4}=\frac{-\sqrt{\pi}}
{32\pi^2\tilde{s}^{3/2}}\nonumber
\end{eqnarray}
where $\tilde{s}=-i(s+i\epsilon)$.
The $<A(\tilde{s})+\tilde{A}(\tilde{s})>$ due to (\ref{coef}) can be written as
\begin{equation}
\label{sum}
<A(\tilde{s})+\tilde{A}(\tilde{s})>=2\sum_{n=0}^{\infty}
\frac{{(4\tilde{s}\sqrt{h\bar{h}\varphi\bar{\varphi}})}^{2n+2}}{(2n+2)!}\Box^n
\end{equation}
After using (\ref{ints}) and (\ref{sum}) and carrying out the Wick rotation 
we obtain the contribution to the effective action corresponding to 
general effective potential in the form
\begin{eqnarray}
K^{(1)}&=&\int d^8 z
\int_0^{\infty}\frac{d\tilde{s}}{\tilde{s}}\sum_{l=0}^{\infty}\big[
\frac{{(4\sqrt{h\bar{h}}\tilde{s})}^{4l+2}}{(4l+2)!}
{(\varphi\bar{\varphi})}^{2l+1}\Box^{2l}U|_{x=x'}+\\
&+&\frac{{(4\sqrt{h\bar{h}}\tilde{s})}^{4l+4}}{(4l+4)!}
{(\varphi\bar{\varphi})}^{2l+2}\Box^{2l+1}U|_{x=x'}\big]\nonumber
\end{eqnarray}
Or in the equivalent form
\begin{eqnarray}
K^{(1)}&=&\int d^8 z\int_0^{\infty} d\tilde{s} \sum_{l=0}^{\infty}\big[
\frac{{(16h\bar{h}\varphi\bar{\varphi}))}^{2l+1}}{(4l+2)!}
\frac{{(-1)}^l l!}{32\pi^2}\tilde{s}^{3l}-\\
&-&\frac{{(16h\bar{h}\varphi\bar{\varphi}))}^{2l+2}}{(4l+4)!}
\frac{{(-1)}^{l+1} (2l+1)!!}{32\pi^{3/2}}\tilde{s}^{3l+3/2}\big]\nonumber
\end{eqnarray}
Now let us put $u={(16h\bar{h}\varphi\bar{\varphi})}^{2/3}\tilde{s}$.
It leads to the expression
\begin{eqnarray}
K^{(1)}&=&\Lambda^{1/3}\bar{\Lambda}^{1/3}{\Big(\frac{{(4\pi)}^2}{Q^2}\Big)}
^{2/3}\int d^8 z e^{\sigma+\bar{\sigma}}\times\\
&\times&\int_0^{\infty} du \sum_{l=0}^{\infty}
\Big(\frac{u^{3l}{(-1)}^l l!}{32\pi^2(4l+2)!}-
\frac{u^{3l+3/2}{(-1)}^l (2l+1)!!}{32\pi^{3/2}2^l(4l+4)!}\Big)\nonumber
\end{eqnarray}
(we have restored $\Lambda$ and $\varphi$ using $h=-36\frac
{\pi^2\Lambda}{Q^2}$,
$\varphi\bar{\varphi}=e^{3(\sigma+\bar{\sigma})}$)
The contribution to general effective action has the form
\begin{equation}
K^{(1)}=\frac{c}{Q^{4/3}}
|\Lambda|^{2/3}\int d^8 z e^{\sigma+\bar{\sigma}}
\end{equation}
where $c$ is a positive finite constant. Namely,
\begin{equation}
\label{kc}
c={(4\pi)}^{4/3}\int_0^{\infty} du \sum_{l=0}^{\infty}
\Big(\frac{u^{3l}{(-1)}^l l!}{32\pi^2(4l+2)!}-
\frac{u^{3l+3/2}{(-1)}^l (2l+1)!!}{32\pi^{3/2}2^l(4l+4)!}\Big)
\end{equation}
The proof of finiteness of this constant is analogous with the proof of
finiteness of constants arisen when calculating chiral effective action which
will be carried out in Appendix B.
\newpage

{\large\bf{Appendix B. Calculation of one-loop chiral effective lagrangian}}

\vspace{2mm}

\setcounter{equation}{0}
\renewcommand{\theequation}{B.\Roman{equation}}

Due to (\ref{cf}) we can represent $A+\tilde{A}$ in the form
\begin{eqnarray}
\label{sm}
A+\tilde{A}&=&
512h\bar{h}\varphi\tilde{s}^2+\sum_{l=0}^{\infty}\Big\{
\frac{{(16^2h\bar{h})}^{l+1}
\varphi^{l+1}}{(2l+2)!}\tilde{s}^{2l+2}+
9\Lambda\frac{{(16^2h\bar{h})}^{l+1}\varphi^l}
{(2l+3)!}\times\\
&\times&\tilde{s}^{2l+3}D^2\varphi
+2^{21}h\bar{h}^2 \frac{D^{\alpha}\varphi 
D_{\alpha}\varphi}{\varphi}
{(16^2h\bar{h})}^l\varphi^l\tilde{s}^{2l+3}
\big(\frac{1}{(2l+2)!}-\frac{3}{(2l+3)!}\big)\Big\}\Box^l\nonumber
\end{eqnarray}

Now we are again use the rule (\ref{kahl}) and (\ref{ints}) and carry out the 
Wick rotation.
Transforming (\ref{kahl}) to the form of the integral over chiral subspace we 
arrive at
\begin{eqnarray}
{\cal L}^{(1)}_c&=&\frac{1}{2} \int_0^{\infty}d\tilde{s}
(-\frac{1}{4}\bar{D}^2)
\sum_{k=0}^{\infty}\Big\{
-\big[4h\bar{h}^2\frac{{(16^2h\bar{h})}^{2k}\varphi^{2k}}{(4k+3)!}
D^2\varphi+
8^3h
\bar{h}^2{(16^2h\bar{h})}^{2k}\varphi^{2k-1}\times\nonumber\\
&\times&
\big(\frac{1}{(4k+2)!}
-\frac{3}{(4k+3)!}\big)D^{\alpha}\varphi D_{\alpha}\varphi\big]
\frac{{(-1)}^k k!}{32\pi^2}\tilde{s}^{3k+1}+\nonumber\\
&+&\big[4h\bar{h}^2\frac{{(16^2h\bar{h})}^{2k}\varphi^{2k}}{(4k+5)!}D^2\varphi+
8^3h\bar{h}^2{(16^2h\bar{h})}^{2k+1}\varphi^{2k+1}\times\nonumber\\
&\times&\big
(\frac{1}{(4k+4)!}
-\frac{3}{(4k+5)!}\big)D^{\alpha}\varphi D_{\alpha}\varphi\big]
\frac{{(-1)}^k (2k+1)!!}{2^k 32\pi^{3/2}}\tilde{s}^{3k+5/2}\big\}\nonumber
\end{eqnarray}
Terms of the form $\bar{D}^2\Phi^n$ are omitted here since they are evidently 
equal to zero.
Let us put $u={(16\sqrt{h\bar{h}})}^{4/3}\varphi^{2/3}\tilde{s}$. Then we obtain
\begin{eqnarray}
\label{res}
{\cal L}^{(1)}_c&=&\frac{1}{2} \int_0^{\infty}du(-\frac{1}{4}\bar{D}^2)
\frac{1}{{(16\sqrt{h\bar{h}})}^{4/3}\varphi^{2/3}}\sum_{k=0}^{\infty}u^{3k}
\Big\{-\big[4h\bar{h}^2\frac{1}{(4k+3)!}\times\\
&\times&D^2\varphi+
8^3h\bar{h}^2\varphi^{-1}
\big(\frac{1}{(4k+2)!}-\frac{3}{(4k+3)!}\big)
D^{\alpha}\varphi D_{\alpha}\varphi\big]\times\nonumber\\
&\times&\frac{{(-1)}^k k!}{32\pi^2}\frac{u}{{(16\sqrt{h\bar{h}})}^{4/3}
\varphi^{2/3}}+
\big[4h\bar{h}^2\frac{1}{(4k+5)!}D^2\varphi+
8^3h\bar{h}^2(16^2h\bar{h})\varphi\times\nonumber\\
&\times&\big(\frac{1}{(4k+4)!}-\frac{3}{(4k+5)!}\big)
D^{\alpha}\varphi D_{\alpha}\varphi\big]
\frac{{(-1)}^k (2k+1)!!}{2^k 32\pi^{3/2}}\frac{u^{5/2}}
{{(16\sqrt{h\bar{h}})}^{10/3}\varphi^{5/3}}\Big\}
\nonumber
\end{eqnarray}

It is equal to
\begin{eqnarray}
\label{res1}
{\cal L}^{(1)}_c&=&2 \int_0^{\infty} du
\sum_{k=0}^{\infty}u^{3k}\big\{
\partial_m\varphi\partial^m\varphi\times\nonumber\\
&\times&\big[8^3h^{-1/3}\bar{h}^{2/3}
\big(\frac{1}{(4k+2)!}-\frac{3}{(4k+3)!}\big)
\frac{{(-1)}^k k!}{32\pi^2}\frac{u}{\varphi^{7/3}}-\nonumber\\
&-&\big(\frac{1}{(4k+4)!}-\frac{3}{(4k+5)!}\big)
\frac{{(-1)}^k (2k+1)!!}{2^k 32\pi^{3/2}}h^{2/3}\bar{h}^{-1/3}
\frac{u^{5/2}}{\varphi^{4/3}}\big]
+\nonumber\\
&+&\Box\varphi
\big[h^{-1/3}\bar{h}^{2/3}
\frac{1}{(4k+3)!}\frac{{(-1)}^k k!}{32\pi^2}\frac{u}{\varphi^{4/3}}-\nonumber\\
&-&h^{-4/3}\bar{h}^{5/3}\frac{1}{(4k+5)!}\frac{{(-1)}^k (2k+1)!!}
{2^k 32\pi^{3/2}}\frac{u^{5/2}}{\varphi^{7/3}}\big]\big\}
\end{eqnarray}

The constants $c_1$, $c_2$, $c_3$, $c_4$ introduced in (\ref{chir}) have the
form
\begin{eqnarray}
\label{kons}
c_1&=&18\int_0^{\infty}du \sum_{k=0}^{\infty}u^{3k+1} 8^3
\big(\frac{1}{(4k+2)!}-\frac{3}{(4k+3)!}\big)
\frac{{(-1)}^k k!}{32\pi^2} \nonumber\\
c_2&=&-18\int_0^{\infty}du \sum_{k=0}^{\infty}u^{3k+5/2}
\big(\frac{1}{(4k+4)!}-\frac{3}{(4k+5)!}\big)
\frac{{(-1)}^k (2k+1)!!}{2^k 32\pi^{3/2}}\\
c_3&=&6\int_0^{\infty}du\sum_{k=0}^{\infty}u^{3k+1}\frac{1}{(4k+3)!}
\frac{{(-1)}^k k!}{32\pi^2} \nonumber\\
c_4&=&-6\int_0^{\infty}du \sum_{k=0}^{\infty}u^{3k+5/2}
\frac{1}{(4k+5)!}\frac{{(-1)}^k (2k+1)!!}{2^k 32\pi^{3/2}}\nonumber
\end{eqnarray}
All these integrals over $u$ are finite. Let us prove this statement.

We start with the following integrals
\begin{eqnarray}
I_1&=&\int_0^{\infty}du \sum_{k=0}^{\infty}u^{3k+1}\frac{{(-1)}^k k!}{(4k+2)!}\\
I_2&=&\int_0^{\infty}du \sum_{k=0}^{\infty}u^{3k+5/2}
\frac{{(-1)}^k (2k+1)!!}{2^k (4k+4)!}\nonumber
\end{eqnarray}
Let us consider the identity
$$\frac{(2n+1)!!}{2^n}=\frac{1}{\sqrt{\pi}}\Gamma(n+\frac{3}{2})=
\frac{1}{\sqrt{\pi}}\int_0^{\infty} dt e^{-t} t^{n+\frac{1}{2}}$$
It leads to
\begin{eqnarray}
I_1&=&\int_0^{\infty}du u\int_0^{\infty} dt e^{-t} {(u^3 t)}^{-1/2}
\sum_{k=0}^{\infty}{({(u^3 t)}^{1/4})}^{4k+2}\frac{{(-1)}^k }{(4k+2)!}\\
I_2&=&\frac{1}{2\sqrt{\pi}}
\int_0^{\infty}du u\int_0^{\infty} dt e^{-t} {(u^3 t)}^{-1/2}
\sum_{k=0}^{\infty}{({(u^3 t)}^{1/4})}^{4k+4}\frac{{(-1)}^k}{(4k+4)!}\nonumber
\end{eqnarray}
Let us use the following formulas
\begin{eqnarray}
\sum_{n=0}^{\infty}\frac{a^{4n+2}}{(4n+2)!}{(-1)}^n&=&
-\frac{i}{2}(\cosh\frac{1+i}{\sqrt{2}}a-\cosh\frac{1-i}{\sqrt{2}}a)\\
\sum_{n=0}^{\infty}\frac{a^{4n+4}}{(4n+4)!}{(-1)}^n&=&
-\frac{1}{2}(\cosh\frac{1+i}{\sqrt{2}}a+\cosh\frac{1-i}{\sqrt{2}}a)+1\nonumber
\end{eqnarray}
As a result we obtain
\begin{eqnarray}
I_1&=&-\frac{i}{2}\int_0^{\infty}du \int_0^{\infty} dt e^{-t} {(u t)}^{-1/2}
(\cosh\frac{1+i}{\sqrt{2}}{(u^3 t)}^{1/4}-\cosh\frac{1-i}{\sqrt{2}}{(u^3 t)}^
{1/4})
\nonumber\\
I_2&=&\frac{1}{2\sqrt{\pi}}
\int_0^{\infty}du \int_0^{\infty} dt e^{-t} {(u t)}^{-1/2}\big(
(-\frac{1}{2})\times\\
&\times&(\cosh\frac{1+i}{\sqrt{2}}{(u^3 t)}^{1/4}+\cosh\frac{1-i}{\sqrt{2}}
{(u^3 t)}^{1/4})+1\big)\nonumber
\end{eqnarray}
These integrals are evidently finite. The analogous consideration shows that 
the other intergals in eq. (\ref{kons}) are also finite.

\end{document}